\documentclass[twocolumn]{article}

\usepackage[pdftex]{graphicx}

\usepackage{cite}
\usepackage{url}
\usepackage{color}
\usepackage{multirow}
\usepackage{amsmath,amssymb}

\renewcommand{\refname}{Reference}
\renewcommand{\figurename}{Fig.}

\begin{document}
\title{\Huge{An agent-based model for designing \\ a financial market that works well}}
\author{\LARGE{Takanobu Mizuta}\thanks{mizutata@gmail.com, http://mizutatakanobu.com} \\ SPARX Asset Management Co. Ltd., Tokyo, Japan}

\date{}

\maketitle

\begin{abstract}

Designing a financial market that works well is very important for developing and maintaining an advanced economy, but is not easy because changing detailed rules, even ones that seem trivial, sometimes causes unexpected large impacts and side effects. A computer simulation using an agent-based model can directly treat and clearly explain such complex systems where micro processes and macro phenomena interact. Many effective agent-based models investigating human behavior have already been developed. Recently, an artificial market model, which is an agent-based model for a financial market, has started to contribute to discussions on rules and regulations of actual financial markets. I introduce an artificial market model to design financial markets that work well and describe a previous study investigating tick size reduction. I hope that more artificial market models will contribute to designing financial markets that work well to further develop and maintain advanced economies.

\end{abstract}



\section{Artificial market model: an agent-based model for a financial market}

\subsection{Importance and difficulty of market design}

People have been able to develop advanced economies by cooperating to exchange goods for money. Creation of any industry requires investment to first purchase or build tools to make goods. Thus, a financial market  that enables smooth investment is obviously required.

The economist John McMillan, who used game theory to investigate many markets, said ``a market works well only if it is well designed''\cite{McMillan2002}. Market design (regulations, rules) determines whether a market works well or badly. McMillan also concluded that ``the economy is a highly complex system. It is at least as complex as the systems studied by physicists and biologists.'' The computer scientist Melanie Mitchell said ``economies are complex systems in which the simple, microscopic components consist of people buying and selling goods, and the collective behavior is the complex, hard-to-predict behavior of markets as a whole, such as fluctuations in stock prices''\cite{mitchell2009complexity}. A financial market is another highly complex system where a simple summation of micro processes (trader behaviors) never explains macro phenomena (price formation). Changing detailed rules, even ones that seem trivial, sometimes causes unexpected large impacts and side effects. McMillan illustrated this nature as ``both God and the devil are in the details.'' Designing a market well is very important for developing an advanced economy, but not easy.

\subsection{An agent-based model explaining a complex system}

Separately investigating macro phenomena and micro processes unclearly explains complex systems where macro phenomena and micro processes interact. A mathematical model and an empirical study cannot directly treat or clearly explain the interactions. A computer simulation using an agent-based model, on the other hand, can directly treat and clearly explain the interactions\footnote{Sabzian et al. provide a comprehensive review of agent-based models for complex systems\cite{Sabzian2018}.}. An agent-based model includes agents modeling trader behaviors and shows macro phenomena as a result of their interactions. Agent behaviors that are simple but affected by macro phenomena cause complex macro phenomena, which are not a simple summation of the agent behaviors. Thus, an agent-based model gives researchers new knowledge. An agent-based model, requiring no data, is a true computer simulation.

Not only financial markets but also social systems are complex system. Many studies using agent-based models has already succeed to investigate social systems . For examples, investigating effects of new creation road ways to traffic jam and determining an evacuation route with terror and fire in a building. To solve such problems, researchers naturally use various approaches: a mathematical model, an empirical study, and an agent-based model. Each approach has advantages and disadvantages and gives researchers various viewpoints and knowledge to find unexpected side effects. An agent-based model for such problems has been as indispensable as a mathematical model and an empirical study.

\subsection{An artificial market model = an agent-based model for a financial market}

An artificial market model is an agent-based model for a financial market. Since the 1990s, many significant artificial market models \cite{izumi1996artificial,arthur1997economy,lux1999scaling} have been developed. Projects building generic artificial market models have been conducted such as the U-mart project in Japan in the 2000s\footnote{Kita et al. provide a comprehensive review \cite{kitarealistic2016}.}. These artificial market models have contributed to explaining the nature of financial market phenomena such as bubbles and crashes.

An artificial market model, however, has rarely been used to investigate the rules and regulations of a financial market. After the bankruptcy of Lehman Brothers in 2008, some articles argued that traditional economics had not found ways to design markets that work well and anticipated an artificial market model to do so. Indeed, in Science, Battiston et al. \cite{Battiston818} explained that ``since the 2008 crisis, there has been increasing interest in using ideas from complexity theory (using network models and agent-based models) to make sense of economic and financial markets,'' and in Nature, Farmer and Foley \cite{farmer2009economy} explained that ``such (agent based) economic models should be able to provide an alternative tool to give insight into how government policies could affect the broad characteristics of economic performance, by quantitatively exploring how the economy is likely to react under different scenarios.'' 

Financial regulators and exchanges, who decide rules and regulations, especially desire an artificial market model to design a market that works well. Indeed, the Japan Exchange Group (JPX), which is the parent company of the Tokyo Stock Exchange, has published 30 JPX working papers including 9 papers using an artificial market model as of April 2019\footnote{\url{https://www.jpx.co.jp/english/corporate/research-study/working-paper/index.html}}.

Also in Europe, a three-year project (2014-2017) founded by the European Commission to integrate macro-financial modeling for robust policy design included a work package named bridging agent-based and dynamic-stochastic-general-equilibrium modeling approaches for building policy-focused macro-financial models\footnote{\url{http://www.macfinrobods.eu/research/workpackages/WP7/wp7.html}}. The Bank of England also published a working paper investigating effects of passive funds in a bond market using an artificial market model \cite{braun2016staff}. 

Mizuta \cite{mizuta2016SSRNrev} reviewed other previous agent-based models for designing a financial market that works well that are not mentioned above.

\section{Suitable complexity, advantages and disadvantages}

Here, I will discuss features that an artificial market model for designing a financial market should have. Such models aim not to accurately forecast but to design a financial market that works well. To discuss what a better design is, acquiring knowledge of what mechanism affects prices is more important than replicating a real financial market. 

Such a model needs to reveal possible mechanisms that affect price formation through many simulation runs, e.g., searching for parameters or purely comparing the before/after of changes. Possible mechanisms revealed by these runs provide new knowledge and insights into the effects of the changes on price formation in actual financial markets. Other methods of study, e.g., empirical studies, would not reveal such possible mechanisms. 

An unnecessary replication of macro phenomena leads to models that are over-fitted and too complex. Such models would prevent us from understanding and discovering mechanisms that affect price formation because the number of related factors would increase. Indeed, artificial market models that are too complex are often criticized because they are very difficult to evaluate\cite{chen2009agent}. A model that is too complex not only would prevent us from understanding mechanisms but also could output arbitrary results by over-fitting too many parameters. It is more difficult for simpler models to obtain arbitrary results, so these models are easier to evaluate. An artificial market model should be built as simple as possible and not intentionally implement agents to cover all the investors who would exist in actual financial markets.

As Michael Weisberg mentioned, modeling is ``the indirect study of real-world systems via the construction and analysis of models. Modeling is not always aimed at purely veridical representation. Rather, the researchers worked hard to identify the features of these systems that were most salient to their investigations.''\cite{Weisberg2012} Therefore, good models differ depending on the phenomena being focused on. Thus, my model is good only for the purpose of this study and may be not good for other purposes. An aim of my study is to understand how important properties (behaviors, algorithms) affect macro phenomena and play a role in the financial system rather than representing actual financial markets precisely.

\begin{figure}[tb]
\begin{center}
  \includegraphics[scale=0.35]{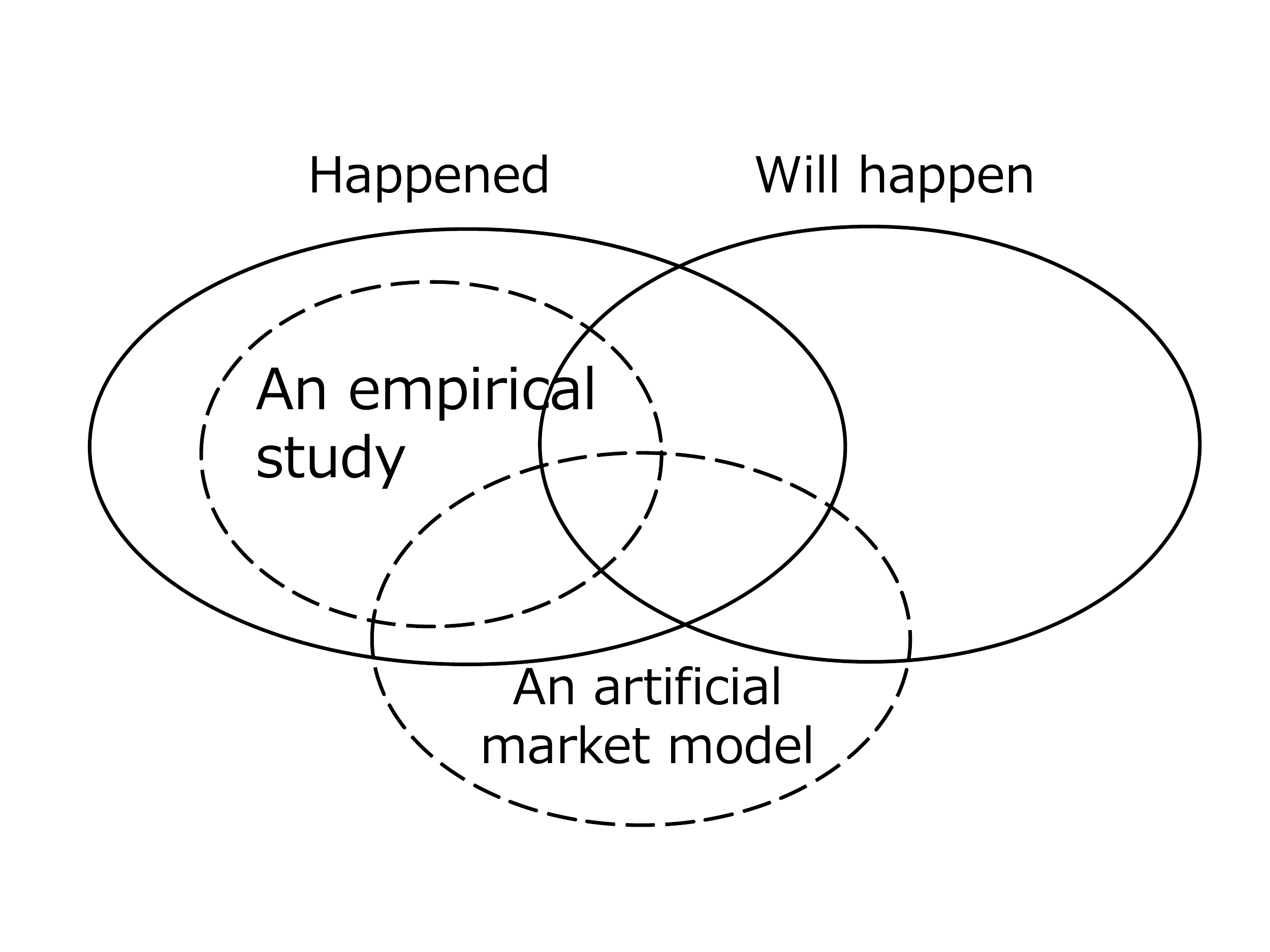}
 \end{center}
  \caption{An artificial market model and an empirical study}
  \label{fig:p4}
\end{figure}

Fig. \ref{fig:p4} shows features of outputs of an artificial market model and an empirical study. Outputs of an empirical study are included in the area that has happened in a real financial market. The advantage of an empirical study is outputs exclude the all area not happening in the past or future. The disadvantage, however, is outputs exclude any area happening in the future. 

The advantage of an artificial market model is outputs include the part of the area happening in the future. The disadvantage, however, is outputs include the part of area not happening in the past or future. An artificial market model just outputs ``possible'' results to understand the mechanism of a market. Discussing whether the results will occur or not needs other methods, e.g., an empirical study and a mathematical model.

Discussing the outputs of an artificial market model always needs knowledge given by empirical studies and mathematical models. A market that works well should be designed by not one but several methods (an artificial market model, empirical study and a mathematical model), and the methods should collaborate to mutually compensate for their disadvantages.

Many empirical studies, e.g., Sewell \cite{Sewell2006}, have shown that both stylized facts (fat-tail and volatility-clustering) exist statistically in almost all financial markets. Conversely, they have also shown that only the fat-tail and volatility-clustering are stably observed for any asset and in any period because financial markets are generally unstable. This leads to the conclusion that an artificial market should replicate macro phenomena existing generally for any asset and any time, fat-tail, and volatility-clustering. This is an example of how empirical studies can help an artificial market model.

\section{Case study: tick size reduction}

\subsection{Tick size reduction}

I introduce a paper investigating tick size reduction \cite{mizuta2013TSEWPe} (vol. 2, JPX working paper) as a typical study investigating the design of a financial market using an artificial market model.

The tick size is the minimum unit of a price change. For example, when the tick size is \$1, order prices such as \$99 and \$100 are accepted, but \$99.1 (\$99.10 cent) is not. Tokyo Stock Exchange used Y\llap{=}1 as the tick size until 18 July 2014 and has used Y\llap{=}0.1 (10 sen) since 22 July 2014.

More stock markets are now making full use of information technology (IT) to achieve low-cost operations, especially in the United States and Europe. Their market shares of trading volume have caught up with those of traditional stock exchanges. Thus, each stock is traded at many stock markets at once. Whether such fragmentation makes markets more efficient has been debated \cite{foucault2008competition,o2011market}. Many factors, such as tick size, speed of trading systems, length of trading hours, stability of trading systems, safety of clearing, and variety of order types determine the market share of trading volume between actual markets. A smallness of the tick size is one of the most important factors to compete with other markets.

Mizuta et al. \cite{mizuta2013TSEWPe} used an artificial market model to investigate competition, in terms of taking market share of trading volume, between two artificial financial markets that have exactly the same specifications except for tick sizes and initial trading volume. 

\subsection{Model}

The model of Chiarella and Iori \cite{chiarella2002simulation} is very simple but replicates long-term statistical characteristics observed in actual financial markets: a fat tail and volatility clustering. In contrast, that of Mizuta et al. \cite{mizuta2013TSEWPe} replicates high-frequency micro structures, such as execution rates, cancel rates, and one-tick volatility, that cannot be replicated with the model of Chiarella and Iori \cite{chiarella2002simulation}. Only fundamental and technical strategies existing generally for any market and any time\footnote{Many empirical studies found these strategies, which are comprehensively reviewed by Menkhoff and Taylor \cite{menkhoff2007obstinate}.} are implemented to the agent model.

\begin{figure}[tb]
 \begin{center}
 \includegraphics[scale=0.35]{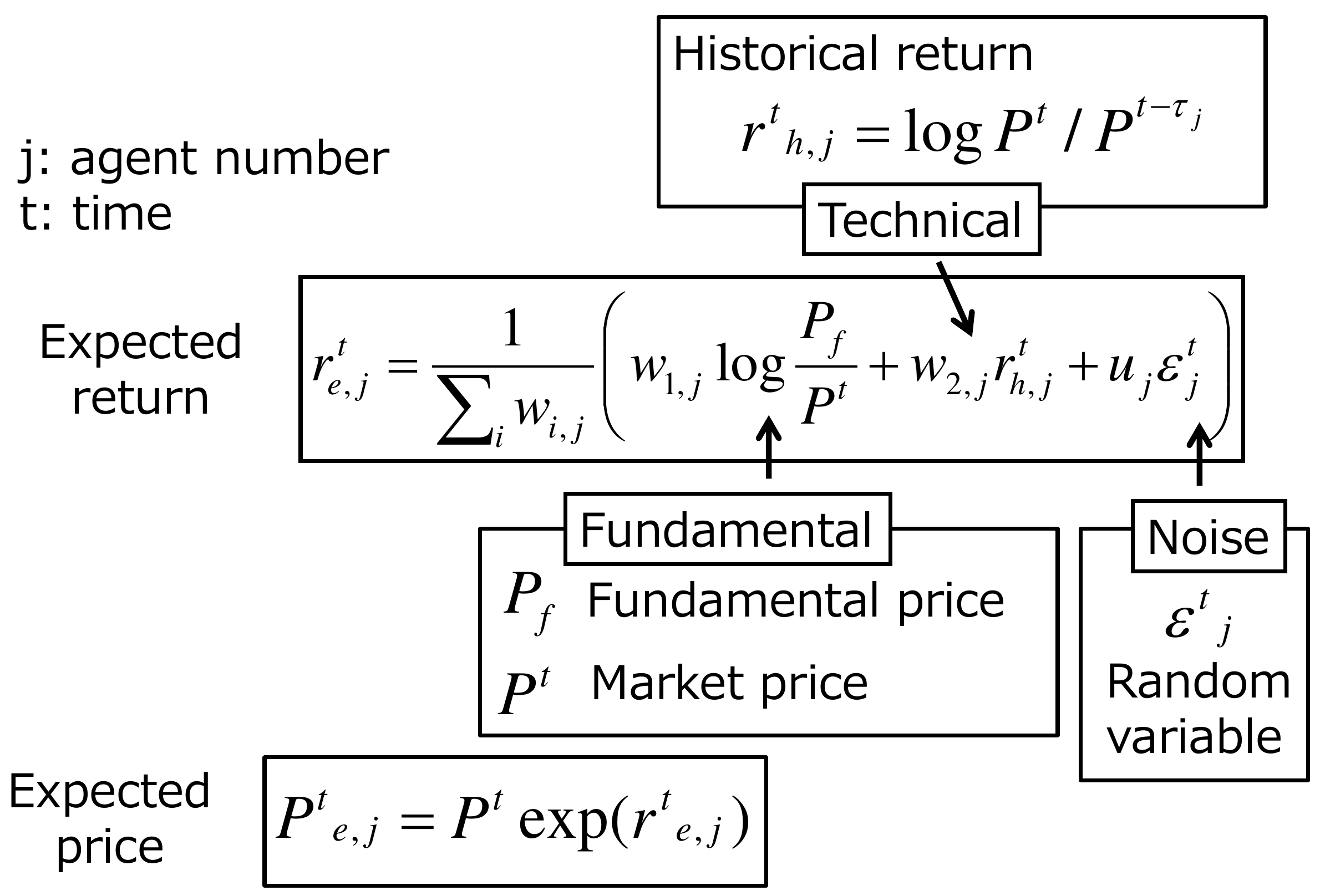}
 \end{center}
  \caption{An agent model}
  \label{fig:p5}
\end{figure}

The number of agents is $n$. First, at time $t=1$, agent $1$ orders to buy or sell the risk asset; then at $t=2$ agent $2$ orders to buy or sell. At $t=3,4,,,n$, agents $3,4,,,n$ respectively order to buy or sell. At $t=n+1$, going back to the first agent, agent $1$ orders to buy or sell, and at $t=n+2,n+3,,n+n$, agents $2,3,,,,n$ respectively order to buy or sell, and this cycle is repeated. Note that $t$ passes even if no deals have occurred. An agent $j$ determines an order price and buys or sells by the following process. Agents use a combination of the fundamental value and technical rules to form expectations on a risk asset return (Fig. \ref{fig:p5}). The expected return of agent $j$ at $t$ is
\begin{equation}
r^{t}_{e,j}=\frac{1}{\Sigma_k^3 w_{k,j}} \left( w_{1,j} \log{\frac{P_f}{P^t}} + w_{2,j}r^t_{h,j}+w_{3,j} \epsilon ^t _j \right), \label{eq0}
\end{equation}
where $w_{i,j}$ is the weight of term $i$ for agent $j$ and is independently determined by random variables uniformly distributed on the interval $(0,w_{i,max})$ at the start of the simulation for each agent. $P_f$ is a fundamental value and is constant\footnote{This enables focusing on phenomena in short time scales, as the fundamental price remains static.}. In addition, $P^t$ is the market price of the risk asset, and $\epsilon ^t _ j$ is determined by random variables from a normal distribution with average $0$ and variance $\sigma _ \epsilon$. Finally, $r^t_{h,j}$ is a historical price return inside an agent's time interval $\tau _ j$, where $r^t_{h,j}=\log{(P^t/P^{t-\tau _ j})}$, and $\tau_j$ is independently determined by random variables uniformly distributed on the interval $(1,\tau _{max})$ at the start of the simulation for each agent\footnote{When $t< \tau _ j$, however, $r^t_{h,j}=0$.}. 

The first term of Eq. (\ref{eq0}) represents a fundamental strategy: the agent expects a positive return when the market price is lower than the fundamental value, and vice versa. The second term of Eq. (\ref{eq0}) represents a technical strategy: the agent expects a positive return when the historical market return is positive, and vice versa. 

After the expected return has been determined, the expected price is
\begin{equation}
P^t_{e,j}= P^t \exp{(r^t_{e,j})}.
\end{equation}

\begin{figure}[tb]
 \begin{center}
  \includegraphics[scale=0.35]{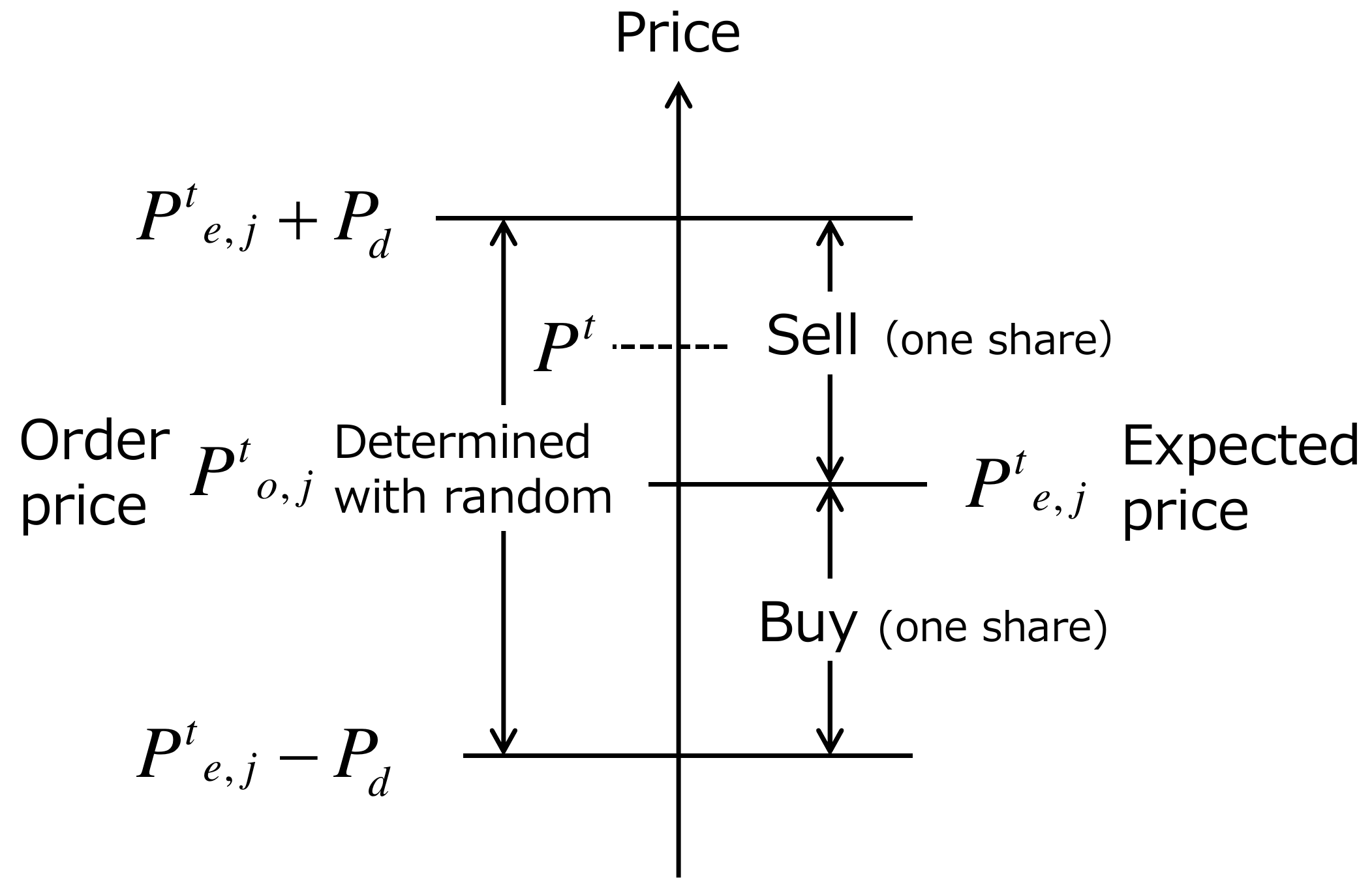}
 \end{center}
  \caption{Scattering the order prices around the expected price.}
  \label{fig:p6}
\end{figure}

The order price $P^t_{o,j}$ is determined by random variables normally distributed with average $P^t_{e,j}$ and standard deviation $P_\sigma$, where $P_\sigma$ is a constant. (Fig. \ref{fig:p6}) Whether to buy or sell is determined by the magnitude relationship between $P^t_{e,j}$ and $P^t_{o,j}$: 

when $P^t_{e,j}>P^t_{o,j}$, the agent places an order to buy one share, but 

when $P^t_{e,j}<P^t_{o,j}$, the agent places an order to sell one share\footnote{When $t<t_c$, however, to generate enough waiting orders, the agent places an order to buy one share when $P_f>P^t_{o,j}$, or to sell one share when $P_f<P^t_{o,j}$.}.

Scattering the order prices around the expected price enables the distribution of order prices in a real financial market to be replicated and a simulation to run stably.

Agents always order only one share. The model adopts a continuous double auction, so when an agent orders to buy (sell), if there is a lower price sell order (a higher price buy order) than the agent's order, dealing immediately occurs. Such an order is called ``market order''. If there is not a lower price sell order (a higher price buy order) than the agent's order, the agent's order remains in the order book. Such an order is called ``limit order''. The remaining order is canceled after $t_c$ from the order time. Agents can short sell freely. The quantity of holding positions is not limited, so agents can take any shares for both long and short positions to infinity. 

\begin{figure}[tb]
 \begin{center}
  \includegraphics[scale=0.35]{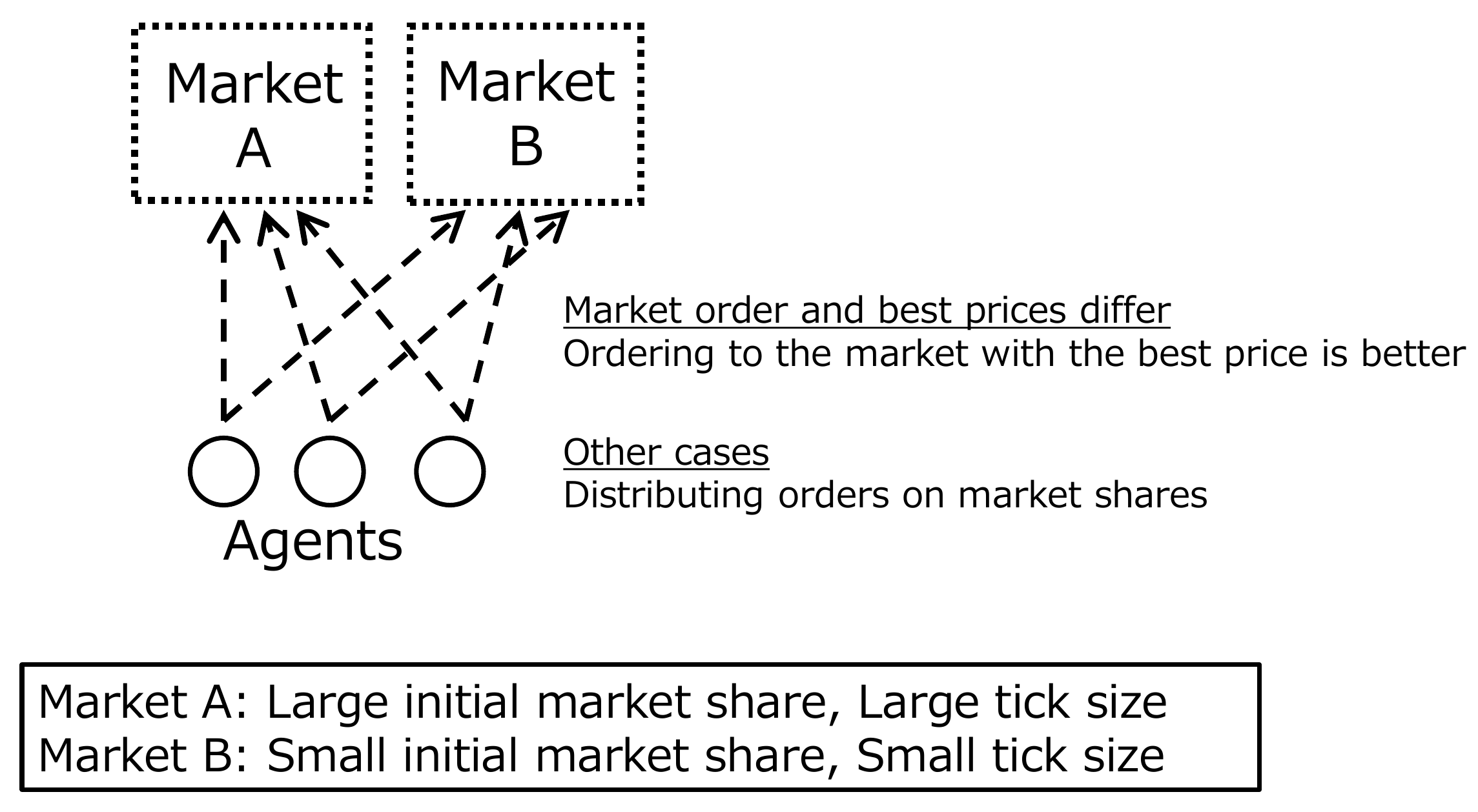}
 \end{center}
  \caption{A model selecting a market to which to order.}
  \label{fig:p9}
\end{figure}

The agents trade one stock at two markets: A and B (Fig. \ref{fig:p9}). The two stock markets have exactly the same specifications except for the minimum unit of a price change (tick size) per $P_f$, $\Delta P_A, \Delta P_B$ and initial share of trading volume, $W_A, W_B$. The agents should decide to which market they order: A or B.

The model of market selection is almost the same as an order allocation algorithm (Smart Order Routing, SOR) used in a real financial market. Each agent determines a market to which to order for every order. When the agent order is buy (sell), the agent searches for the lowest sell (highest buy) orders of each market. These prices are called ``best prices.'' When best prices differ between two markets and the order will be a market order in least one of the markets, the agent orders to buy (sell) in a market in which the best price is better, {\it i.e.,} lower (higher) in the case of the buy (sell) order. In other cases, {\it i.e.,} when the best prices are exactly the same or the order will be a limit order in both markets, the agent orders to buy (sell) in market A with probability $W_A$，

\begin{equation}
W_A=\frac{T_A}{T_A+T_B},
\end{equation}
where $T_A$ is the trading volume of market A within last $t_{AB}$, and the calculating span of $W_A$ and $T_B$ is that of market B. To summarize, if the market order and best prices differ, agents order to buy (sell) in the market in which the best price is better than that in the other market. In other cases, agents order to buy (sell) in markets depending on the market share of trading volume.

\begin{figure}[tb]
 \begin{center}
  \includegraphics[scale=0.35]{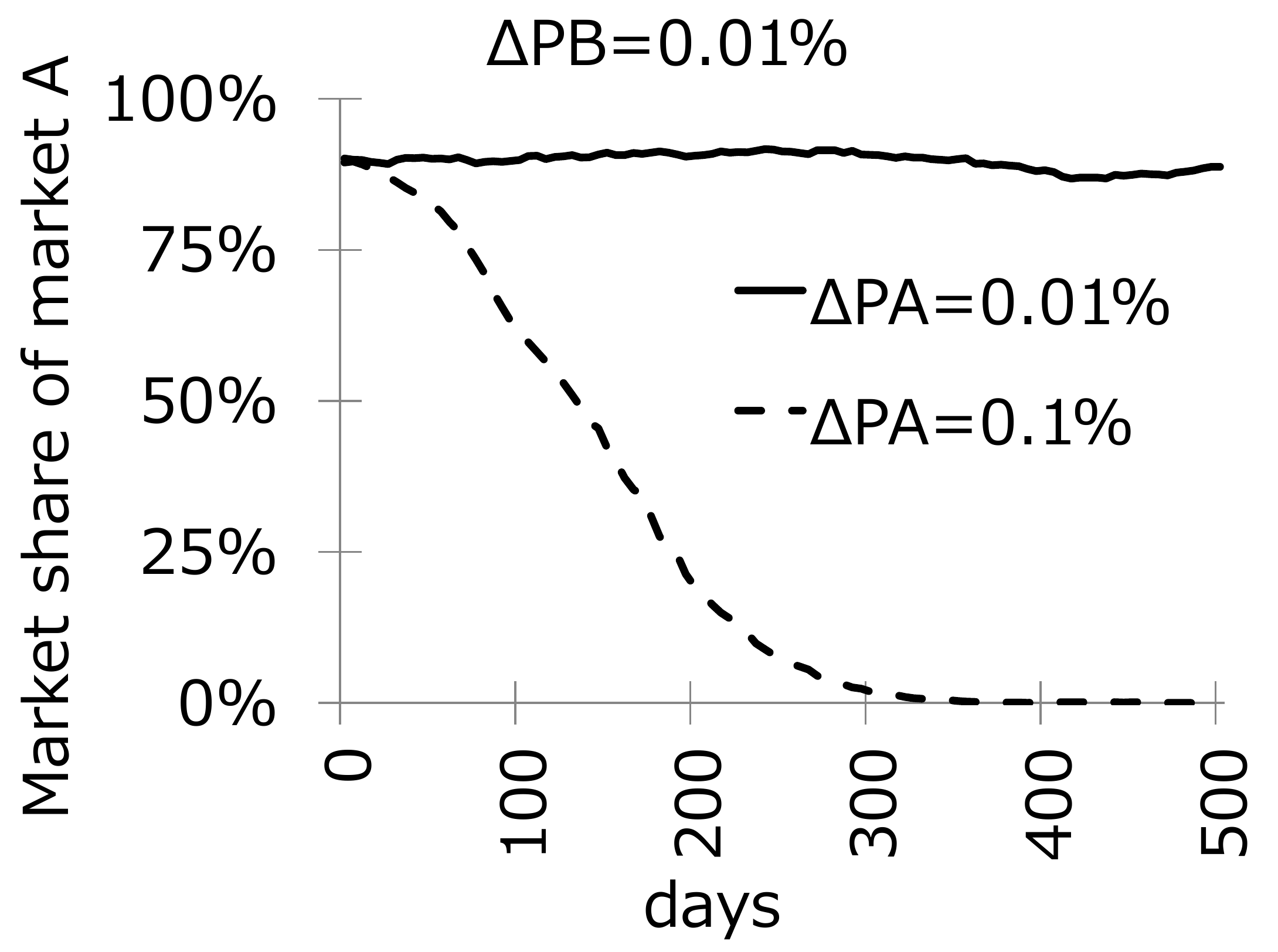}
 \end{center}
  \caption{The time evolution of market shares of trading volume for tick sizes that are not too small.}
  \label{fig:p11}
\end{figure}

\begin{figure}[tb]
 \begin{center}
  \includegraphics[scale=0.35]{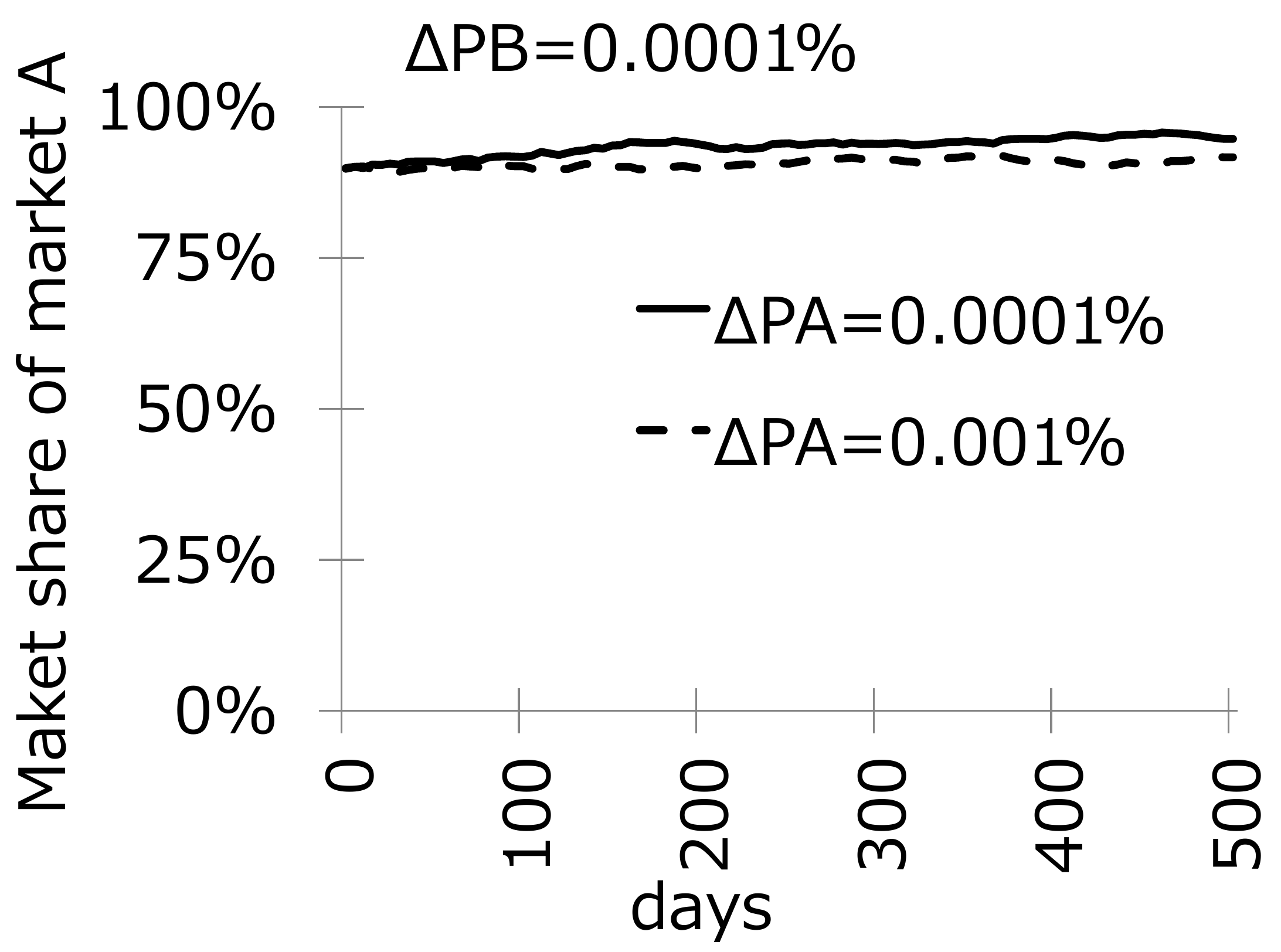}
 \end{center}
  \caption{The time evolution of market shares of trading volume for tick sizes that are too small.}
  \label{fig:p12}
\end{figure}

\subsection{Simulation Results}
Mizuta et al. \cite{mizuta2013TSEWPe} investigated the transition of market shares of trading volume involving two markets. The two stock markets (A and B) had exactly the same specifications except for tick sizes per $P_f$, $\Delta P_A, \Delta P_B$ and initial market share of trading volume $W_A=0.9, W_B=0.1$. They set $n=1000, w_{1,max}=1,w_{2,max}=10,w_{3,max}=1,\tau _ {max}=10000, \sigma _ \epsilon = 0.06, P_\sigma=30, t_c=20000, P_f=10000$, and $t_{AB}=10000$ (5 days). Simulations are ran to $t=10000000$. 

Fig. \ref{fig:p11} shows the time evolution of market share of trading volume of market A, where $\Delta P_A=0.1\%, 0.01\%$ and $\Delta P_B=0.01\%$. In the $\Delta P_A = \Delta P_B = 0.01$ the market shares slightly moved. In $\Delta P_A=0.1$, which is 10 times greater than $\Delta P_B=0.01$, market B took market share of trading volume from market A.

On the other hand, Fig. \ref{fig:p12} shows the case in which $\Delta P_A=0.001\%, 0.0001\%$ and $\Delta P_B=0.0001\%$, which is 1/100 of that in Fig. \ref{fig:p11}, and $\Delta P_A$ also became 1/100. Market B could not take market share despite $\Delta P_B$ being 1/10 of $\Delta P_A$. Therefore, competition under tick sizes that are too small does not affect the taking of market share of trading volume.

\begin{figure}[tb]
 \begin{center}
  \includegraphics[scale=0.35]{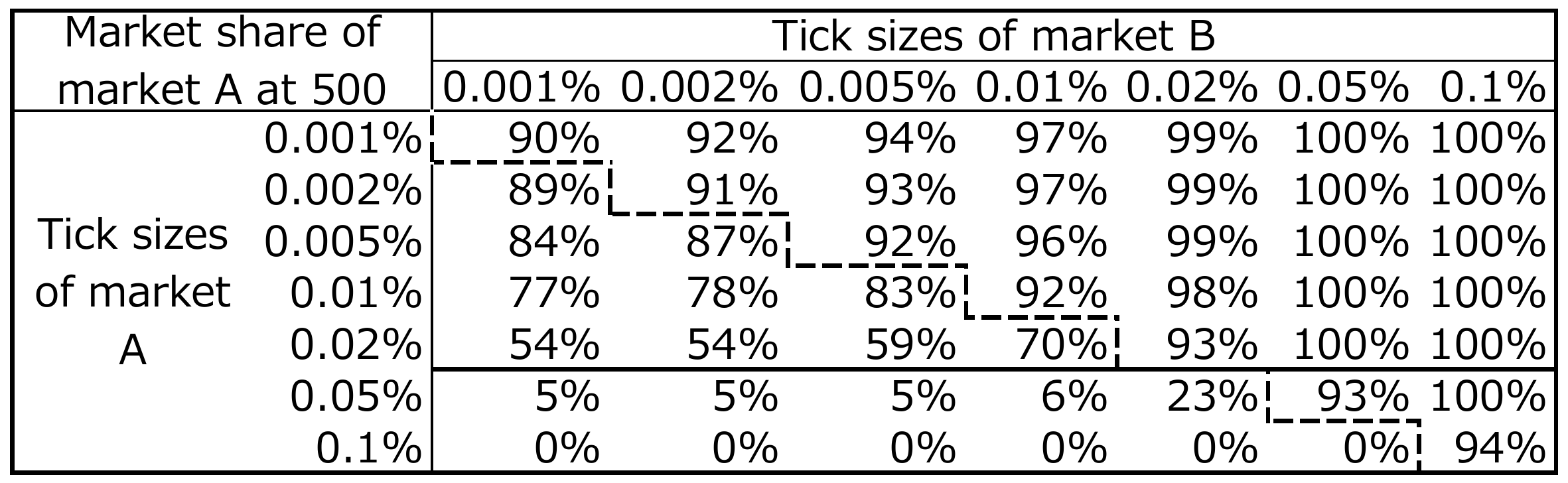}
 \end{center}
  \caption{The market shares of trading volume for various $\Delta P_A$ and $\Delta P_B$.}
  \label{fig:p13}
\end{figure}

The relationship between tick size and taking market share of trading volume is investigated. Fig. \ref{fig:p13} shows the market shares of trading volume of $W_A$ at 500 days for various $\Delta P_A$ and $\Delta P_B$. The following two borderlines, 
\begin{eqnarray}
\Delta P_A \leq \Delta P_B \ {\rm (dashed \ line)}, \label{eq5} \\
\Delta P_A < \overline{\sigma _t} \simeq 0.05\% \ {\rm (solid \ line)}, \label{eq6}
\end{eqnarray}
are drawn where $\overline{\sigma _t}$ is the standard deviation of return for one tick, which was small enough. 

In the region in which at least Eq. (\ref{eq5}) or (\ref{eq6}) is satisfied, market share of trading volume of market A is rarely taken. In the region in which neither Eq. (\ref{eq5}) nor (\ref{eq6}) is satisfied, under the dashed and solid lines, market share of trading volume of market A is rapidly taken. This shows that when the tick size of market A is smaller than $\overline{\sigma _t}$, market share of trading volume of market A is rarely taken even if the tick size of market B is much smaller than that of A.

\begin{figure}[tb]
 \begin{center}
  \includegraphics[scale=0.35]{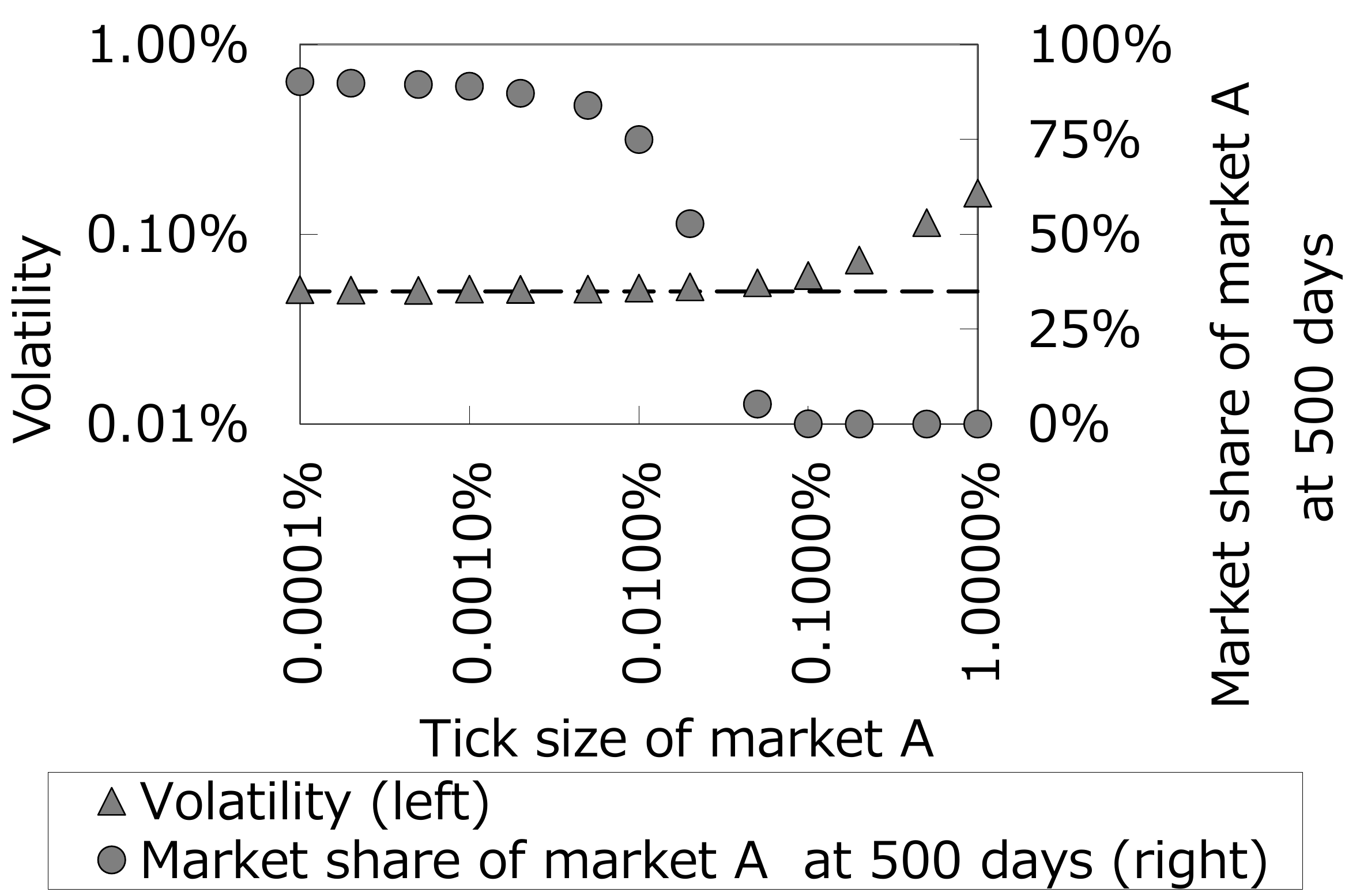}
 \end{center}
  \caption{Tick sizes, volatility and market share of trading volume.}
  \label{fig:p14}
\end{figure}

Fig. \ref{fig:p14} shows the standard deviation of return for one tick (volatility), $\sigma _t$, and market share of trading volume of market A at 500 days, $W_A$ for various $\Delta P_A$ where $\Delta P_B=0.0001\%$. The left vertical axis and horizontal axis are logarithmic scales. Eq. (\ref{eq5}) was never satisfied for all $\Delta P_A$ in Fig. \ref{fig:p14}. The horizontal dotted line is $\sigma _t=\overline{\sigma _t}$. On the left side, $\sigma _t$ equals $\overline{\sigma _t}$ and $\sigma _t$ does not depend on $\Delta P_A$. This means that the difference in tick size does not affect price formations where tick sizes are smaller than $\overline{\sigma _t}$. 

On the right side, $\Delta P_A$ is larger, $\sigma _t$ is larger. This implies that the prices normally fluctuate less than $\Delta P_A$; however, price variation less than $\Delta P_A$ is not permitted. Thus, price fluctuations depend on $\Delta P_A$. In this case, market share of trading volume rapidly deceases in accordance with increasing $\Delta P_A$. On the left side, however, the shares are stable.

\begin{figure}[tb]
 \begin{center}
  \includegraphics[scale=0.35]{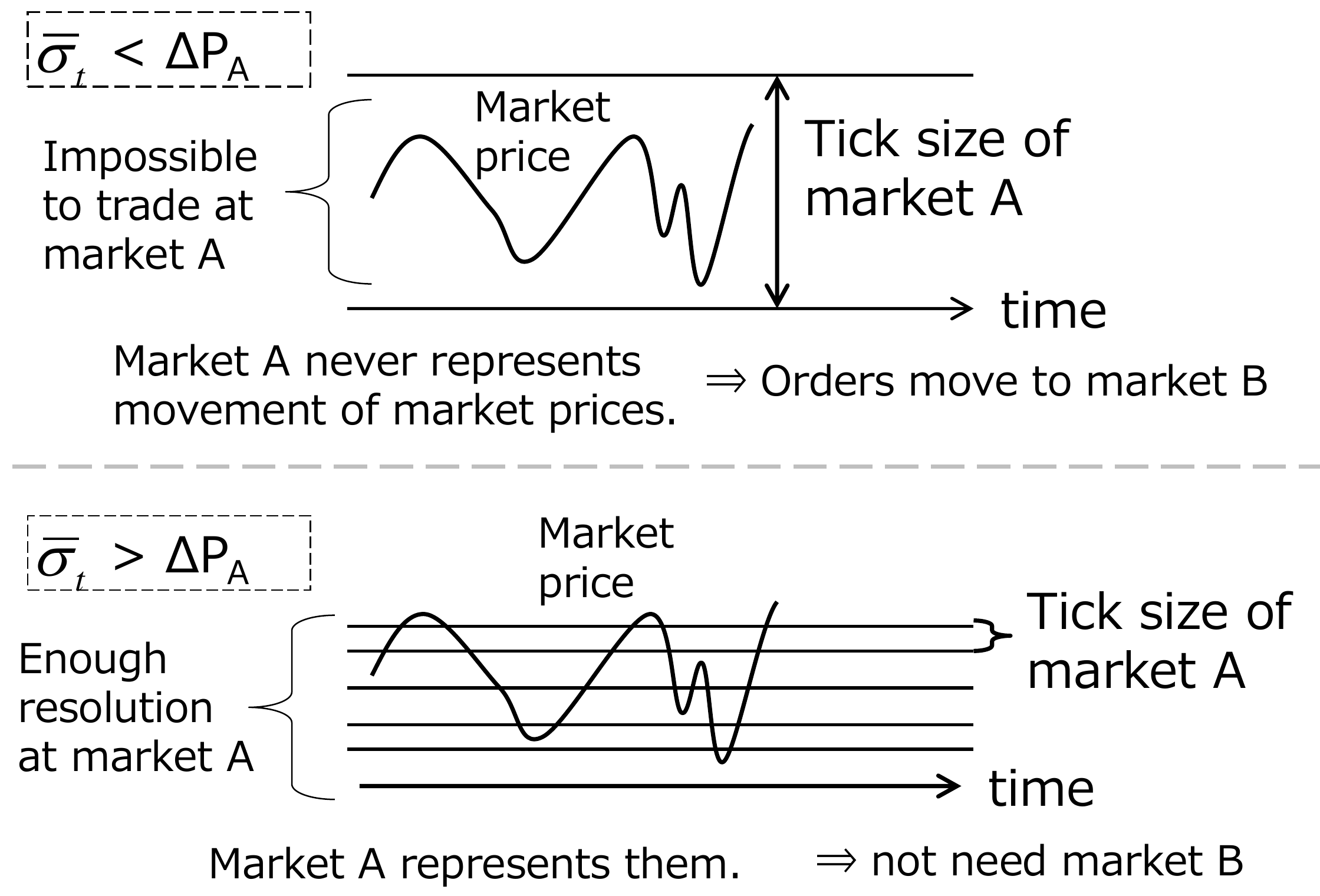}
 \end{center}
  \caption{Mechanism taking market share of trading volume.}
  \label{fig:p15}
\end{figure}

Fig. \ref{fig:p15} summarizes the above discussion. When $\Delta P_A$ is larger than $\overline{\sigma _t}$ (Fig. \ref{fig:p15} top), if $\Delta P_B$ is smaller than $\Delta P_A$, there is a large amount of trading in market B inside $\Delta P_A$. Thus, market B takes market share of trading volume from market A. When $\Delta P_A$ is smaller than $\overline{\sigma _t}$ (Fig. \ref{fig:p15} bottom), even if $\Delta P_B$ is very small, price fluctuations cross many widths of $\Delta P_A$ and sufficient price formations occur only in market A. Thus, market B can rarely take market share of trading volume from market A. 

\begin{figure}[tb]
 \begin{center}
  \includegraphics[scale=0.35]{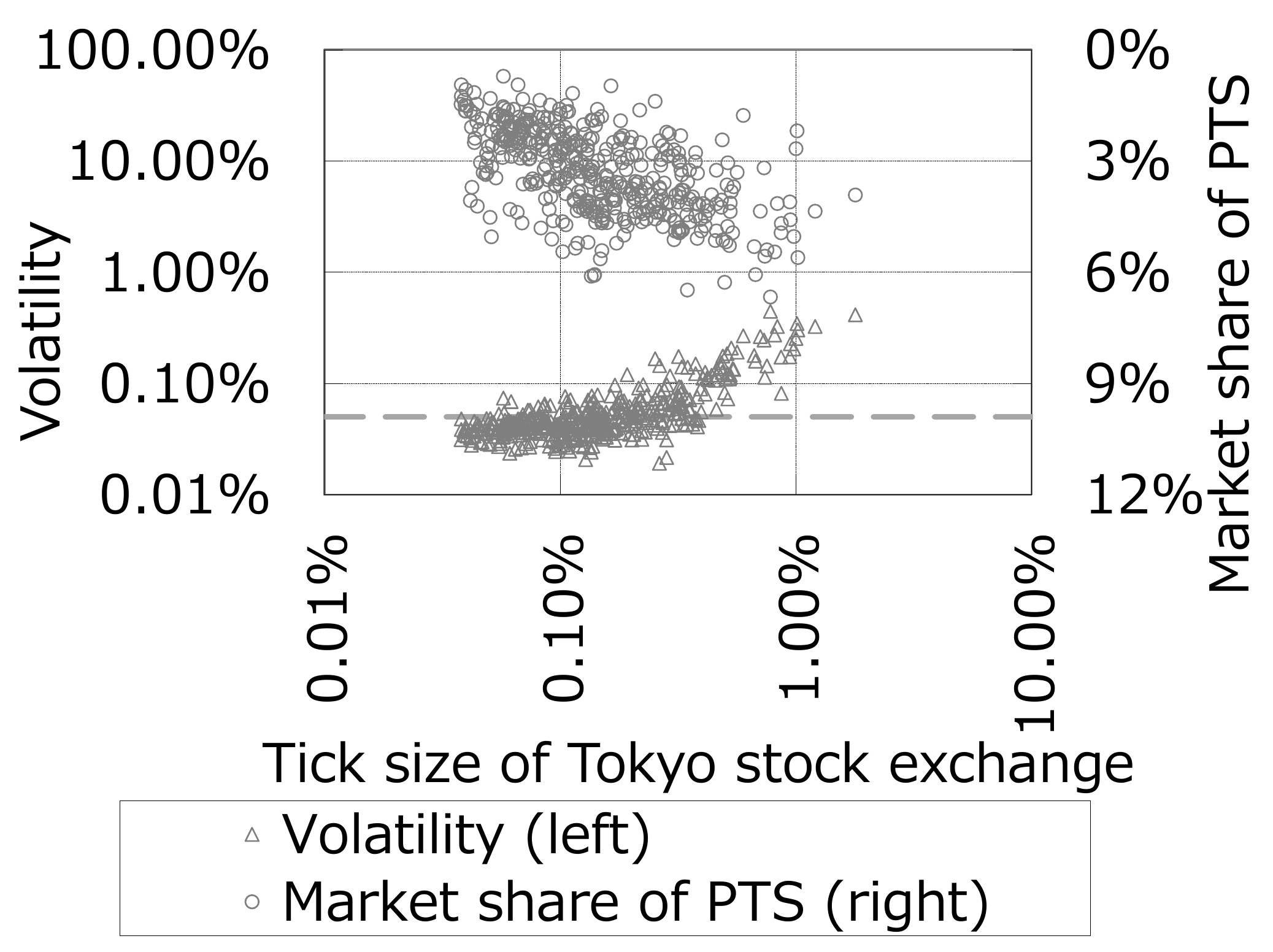}
 \end{center}
  \caption{An empirical analysis for tick sizes, volatility, and market share of trading volume.}
  \label{fig:p16}
\end{figure}

\subsection{An empirical analysis comparing with the simulation results}

Next, Mizuta et al. \cite{mizuta2013TSEWPe} analyzed empirical data and compared them with the simulation results shown in Fig. \ref{fig:p15}, using Japanese stock market data. The data period included all business days in the 2012 calendar year. The number of stocks analyzed is 439, which were selected by TOPIX 500 \footnote{TOPIX 500 is a free-float capitalization-weighted index that is calculated on the basis of the 500 most liquid and highly market capitalized domestic common stocks listed on the Tokyo Stock Exchange first section.} over the entire index data period, had the same minimum unit of a price change for every month end, and were traded every business day. 

Fig. \ref{fig:p16} shows the standard deviations for 10 seconds of each stock (volatility), $\sigma _t$ (triangles), which is the averaged standard deviation of return for 10 minutes except for opening prices for every day, and $\Delta P$ is market share of the trading volume of the Proprietary Trading System (PTS) \footnote{Electric trading systems outside stock exchanges are called PTSs in Japan. A PTS is very similar to an Alternative Trading System (ATS) and Electronic Communications Network (ECN) in other countries.} of each stock (circles) with tick sizes per averaged prices at the end of every month of the Tokyo Stock Exchange\footnote{Mizuta et al. used the data from the Tokyo Stock Exchange to calculate $\Delta P$ and $\sigma _t$. They used Bloomberg data to calculate the market share or trading volume of the PTS, which is its entire trading volume divided by those of Japan’s traditional stock exchanges and PTS, where PTSs are Japan Next PTS J-Market, Japan Next PTS X-Market, and Chi-X Japan PTS, and where Japan’s traditional stock exchanges are the Tokyo, Osaka, Nagoya, Fukuoka, and Sapporo stock exchanges and JASDAQ.}. The right vertical axis is upside down to easily compare with Fig. \ref{fig:p15}. The horizontal dotted line is $\sigma _t=\overline{\sigma _t}$. 

On the left side, $\sigma _t$, which equaled $\overline{\sigma _t}$, did not much depend on $\Delta P$. On the right side, $\Delta P$ and $\sigma _t$ were larger. These results are similar to those in Fig. \ref{fig:p15}. The market share of trading volume of PTS deceased along with $\Delta P$. When $\Delta P$ was larger, PTS more easily took market share of trading volume, and $\sigma _t$ tended to increase along with $\Delta P$.

\subsection{Summary of the case study: tick size reduction}

A market having a tick size larger than volatility will lose market share of trading volume to other markets. In contrast, a market having a tick size smaller than volatility, even if the tick size is larger than those of other markets, will rarely lose market share to other markets. A tick size smaller than volatility rarely affects competition of market share between financial markets, whereas a tick size larger than volatility enlarges the volatility and prevents adequate price formation.

This simulation study is the first to discuss an adequate tick size\footnote{Darley and Outkin \cite{darley2007nasdaq} investigated tick size reduction using an artificial market model when NASDAQ, a stock exchange in the U.S., was planning tick size reduction. They showed, for example, a market can be unstable when some investment strategies increase. However, their model had too many parameters and they focused on which investors earned more, which prevented them from discussing a good design of a financial market.}. An empirical study cannot investigate tick sizes that have never been used in an actual financial market or isolate a direct effect on price formation affected by many factors. In contrast, an artificial market model can isolate the pure contributions of changing a tick size to price formation and simulate a tick size that has never been used. An artificial market model has these advantages over an empirical study.

After the simulation study of Mizuta et al. \cite{mizuta2013TSEWPe}, the mathematical model of Nagumo et al. \cite{doi:10.7566/JPSJ.86.014801} achieved the same result. In this way, an artificial market model can indicate new problems that studies using mathematical models and empirical analysis should solve.

\section{Conclusion}

In this paper, I introduced an artificial market model, which is an agent-based model for a financial market, to design a financial market that works well. An artificial market model has recently started to contribute to discussions on rules and regulations of actual financial markets such as tick size reduction. The contribution has not been great yet but will become greater soon.

Some readers may think tick size reduction is a trivial matter for a financial market. This is, however, important and should not be underestimated. Changing detailed rules sometimes causes unexpected large impacts and side effects. John McMillan illustrated this nature as ``both God and the devil are in the details''\cite{McMillan2002}. Detailed design can determine whether a financial market develops or destroys an advanced economy. Designing a market well is very important for developing and maintaining an advanced economy, but not easy. 

I hope that more artificial market models will contribute to designing a financial market that works well to further develop and maintain advanced economies.

\subsection*{Disclaimer}
\footnotesize{Note that the opinions contained herein are solely those of the authors and do not necessarily reflect those of SPARX Asset Management Co., Ltd.}

\bibliographystyle{IEEEtran}
\bibliography{ref}

\begin{thebibliography}{10}
\providecommand{\url}[1]{#1}
\csname url@samestyle\endcsname
\providecommand{\newblock}{\relax}
\providecommand{\bibinfo}[2]{#2}
\providecommand{\BIBentrySTDinterwordspacing}{\spaceskip=0pt\relax}
\providecommand{\BIBentryALTinterwordstretchfactor}{4}
\providecommand{\BIBentryALTinterwordspacing}{\spaceskip=\fontdimen2\font plus
\BIBentryALTinterwordstretchfactor\fontdimen3\font minus
  \fontdimen4\font\relax}
\providecommand{\BIBforeignlanguage}[2]{{%
\expandafter\ifx\csname l@#1\endcsname\relax
\typeout{** WARNING: IEEEtran.bst: No hyphenation pattern has been}%
\typeout{** loaded for the language `#1'. Using the pattern for}%
\typeout{** the default language instead.}%
\else
\language=\csname l@#1\endcsname
\fi
#2}}
\providecommand{\BIBdecl}{\relax}
\BIBdecl

\bibitem{McMillan2002}
\BIBentryALTinterwordspacing
J.~McMillan, \emph{Reinventing the bazaar: A natural history of markets}.\hskip
  1em plus 0.5em minus 0.4em\relax W. W. Norton \& Company, 2002. [Online].
  Available: \url{https://books.wwnorton.com/books/Reinventing-the-Bazaar/}
\BIBentrySTDinterwordspacing

\bibitem{mitchell2009complexity}
\BIBentryALTinterwordspacing
M.~Mitchell, \emph{Complexity: A guided tour}.\hskip 1em plus 0.5em minus
  0.4em\relax Oxford University Press, 2009. [Online]. Available:
  \url{https://global.oup.com/academic/product/complexity-9780195124415}
\BIBentrySTDinterwordspacing

\bibitem{Sabzian2018}
\BIBentryALTinterwordspacing
H.~Sabzian, M.~A. Shafia, A.~Bonyadi~Naeini, G.~Jandaghi, and M.~J. Sheikh, ``A
  review of agent-based modeling (abm) concepts and some of its main
  applications in management science,'' \emph{Iranian Journal of Management
  Studies}, vol.~11, no.~4, pp. 659--692, 2018. [Online]. Available:
  \url{http://dx.doi.org/10.22059/ijms.2018.261178.673190}
\BIBentrySTDinterwordspacing

\bibitem{izumi1996artificial}
\BIBentryALTinterwordspacing
K.~Izumi and T.~Okatsu, ``An artificial market analysis of exchange rate
  dynamics,'' \emph{Evolutionary Programming V, Proceedings of the Fifth Annual
  Conference on Evolutionary Programming}, pp. 27--36, 1996. [Online].
  Available:
  \url{http://citeseerx.ist.psu.edu/viewdoc/summary?doi=10.1.1.100.3293}
\BIBentrySTDinterwordspacing

\bibitem{arthur1997economy}
\BIBentryALTinterwordspacing
W.~Arthur, S.~Durlauf, D.~Lane, and S.~E. Program, ``Asset pricing under
  endogenous expectations in an arti^^ef^^ac^^81cial stock market,'' \emph{The
  economy as an evolving complex system II}, pp. 15--44, 1997. [Online].
  Available: \url{https://www.crcpress.com/p/book/9780201328233}
\BIBentrySTDinterwordspacing

\bibitem{lux1999scaling}
\BIBentryALTinterwordspacing
T.~Lux and M.~Marchesi, ``Scaling and criticality in a stochastic multi-agent
  model of a financial market,'' \emph{Nature}, vol. 397, no. February, pp.
  498--500, 1999. [Online]. Available: \url{https://doi.org/10.1038/17290 DO}
\BIBentrySTDinterwordspacing

\bibitem{kitarealistic2016}
\BIBentryALTinterwordspacing
H.~Kita, K.~Taniguchi, and Y.~Nakajima, \emph{Realistic Simulation of Financial
  Markets}.\hskip 1em plus 0.5em minus 0.4em\relax Springer, 2016. [Online].
  Available: \url{https://dx.doi.org/10.1007/978-4-431-55057-0}
\BIBentrySTDinterwordspacing

\bibitem{Battiston818}
\BIBentryALTinterwordspacing
S.~Battiston, J.~D. Farmer, A.~Flache, D.~Garlaschelli, A.~G. Haldane,
  H.~Heesterbeek, C.~Hommes, C.~Jaeger, R.~May, and M.~Scheffer, ``Complexity
  theory and financial regulation,'' \emph{Science}, vol. 351, no. 6275, pp.
  818--819, 2016. [Online]. Available:
  \url{http://science.sciencemag.org/content/351/6275/818}
\BIBentrySTDinterwordspacing

\bibitem{farmer2009economy}
\BIBentryALTinterwordspacing
J.~D. Farmer and D.~Foley, ``The economy needs agent-based modelling,''
  \emph{Nature}, vol. 460, no. 7256, pp. 685--686, 2009. [Online]. Available:
  \url{https://www.nature.com/articles/460685a}
\BIBentrySTDinterwordspacing

\bibitem{braun2016staff}
\BIBentryALTinterwordspacing
K.~Braun-Munzinger, Z.~Liu, and A.~Turrell, ``Staff working paper no. 592 an
  agent-based model of dynamics in corporate bond trading,'' \emph{Bank of
  England, Staff Working Papers}, 2016. [Online]. Available:
  \url{http://www.bankofengland.co.uk/research/Pages/workingpapers/2016/swp592.aspx}
\BIBentrySTDinterwordspacing

\bibitem{mizuta2016SSRNrev}
\BIBentryALTinterwordspacing
T.~Mizuta, ``A brief review of recent artificial market simulation (agent-based
  model) studies for financial market regulations and/or rules,'' \emph{SSRN
  Working Paper Series}, 2016. [Online]. Available:
  \url{http://ssrn.com/abstract=2710495}
\BIBentrySTDinterwordspacing

\bibitem{chen2009agent}
\BIBentryALTinterwordspacing
S.-H. Chen, C.-L. Chang, and Y.-R. Du, ``Agent-based economic models and
  econometrics,'' \emph{Knowledge Engineering Review}, vol.~27, no.~2, pp.
  187--219, 2012. [Online]. Available:
  \url{http://dx.doi.org/10.1017/S0269888912000136}
\BIBentrySTDinterwordspacing

\bibitem{Weisberg2012}
\BIBentryALTinterwordspacing
M.~Weisberg, \emph{Simulation and Similarity: Using Models to Understand the
  World}.\hskip 1em plus 0.5em minus 0.4em\relax Oxford Studies in the
  Philosophy of Science, 2012. [Online]. Available:
  \url{https://dx.doi.org/10.1093/acprof:oso/9780199933662.001.0001}
\BIBentrySTDinterwordspacing

\bibitem{Sewell2006}
\BIBentryALTinterwordspacing
M.~Sewell, ``Characterization of financial time series,'' \emph{Research Note,
  University College London, Department of Computer Science}, no. RN/11/01,
  2011. [Online]. Available:
  \url{http://finance.martinsewell.com/stylized-facts/}
\BIBentrySTDinterwordspacing

\bibitem{mizuta2013TSEWPe}
\BIBentryALTinterwordspacing
T.~Mizuta, S.~Hayakawa, K.~Izumi, and S.~Yoshimura, ``Investigation of
  relationship between tick size and trading volume of markets using artificial
  market simulations,'' in \emph{JPX working paper}, no.~2.\hskip 1em plus
  0.5em minus 0.4em\relax Japan Excgange Group, 2013. [Online]. Available:
  \url{https://www.jpx.co.jp/english/corporate/research-study/working-paper/index.html}
\BIBentrySTDinterwordspacing

\bibitem{foucault2008competition}
\BIBentryALTinterwordspacing
T.~Foucault and A.~J. Menkveld, ``Competition for order flow and smart order
  routing systems,'' \emph{The Journal of Finance}, vol.~63, no.~1, pp.
  119--158, 2008. [Online]. Available:
  \url{https://doi.org/10.1111/j.1540-6261.2008.01312.x}
\BIBentrySTDinterwordspacing

\bibitem{o2011market}
\BIBentryALTinterwordspacing
M.~O'Hara and M.~Ye, ``Is market fragmentation harming market quality?''
  \emph{Journal of Financial Economics}, vol. 100, no.~3, pp. 459--474, 2011.
  [Online]. Available: \url{https://doi.org/10.1016/j.jfineco.2011.02.006}
\BIBentrySTDinterwordspacing

\bibitem{chiarella2002simulation}
\BIBentryALTinterwordspacing
C.~Chiarella and G.~Iori, ``A simulation analysis of the microstructure of
  double auction markets,'' \emph{Quantitative Finance}, vol.~2, no.~5, pp.
  346--353, 2002. [Online]. Available:
  \url{https://doi.org/10.1088/1469-7688/2/5/303}
\BIBentrySTDinterwordspacing

\bibitem{menkhoff2007obstinate}
\BIBentryALTinterwordspacing
L.~Menkhoff and M.~P. Taylor, ``The obstinate passion of foreign exchange
  professionals: technical analysis,'' \emph{Journal of Economic Literature},
  pp. 936--972, 2007. [Online]. Available:
  \url{http://www.jstor.org/stable/27646888}
\BIBentrySTDinterwordspacing

\bibitem{darley2007nasdaq}
\BIBentryALTinterwordspacing
V.~Darley and A.~V. Outkin, \emph{Nasdaq Market Simulation: Insights on a Major
  Market from the Science of Complex Adaptive Systems}.\hskip 1em plus 0.5em
  minus 0.4em\relax World Scientific Publishing Co., Inc., 2007. [Online].
  Available: \url{https://www.worldscientific.com/worldscibooks/10.1142/6217}
\BIBentrySTDinterwordspacing

\bibitem{doi:10.7566/JPSJ.86.014801}
\BIBentryALTinterwordspacing
S.~Nagumo, T.~Shimada, N.~Yoshioka, and N.~Ito, ``The effect of tick size on
  trading volume share in two competing stock markets,'' \emph{Journal of the
  Physical Society of Japan}, vol.~86, no.~1, p. 014801, 2017. [Online].
  Available: \url{https://doi.org/10.7566/JPSJ.86.014801}
\BIBentrySTDinterwordspacing

\end{thebibliography}

\end{document}